\documentclass[preprint2]{aastex}

\newcommand{\etal}{\mbox{et al.\ }}
\newcommand{\HII}{H\,{\sc ii}}
\newcommand{\sfrm}{\mbox{SF$_{\mbox{\sc rm}}$}}
\newcommand{\sfem}{\mbox{SF$_{\mbox{\sc em}}$}}

\usepackage{psfig}

\shorttitle{Galactic Magnetic Fields and Ionized Gas}
\shortauthors{Haverkorn et al.}

\begin{document}

\title{Magnetic Fields and Ionized Gas in the inner Galaxy: An Outer
  Scale for Turbulence and the Possible Role of \HII\ Regions}

\author{M. Haverkorn\altaffilmark{1}, B. M. Gaensler\altaffilmark{1},
  N.~M. McClure-Griffiths\altaffilmark{2,3},
  John~M. Dickey\altaffilmark{4}, and A.~J. Green\altaffilmark{5}}

\altaffiltext{1}{Harvard-Smithsonian Center for Astrophysics, 60
  Garden Street, Cambridge MA 01238; mhaverkorn@cfa.harvard.edu,
  bgaensler@cfa.harvard.edu}
\altaffiltext{2}{Australia Telescope National Facility, CSIRO,
  P.O. Box 76, Epping, NSW 1710, Australia;
  naomi.mcclure-griffiths@csiro.au} 
\altaffiltext{3}{Bolton Fellow}
\altaffiltext{4}{Department of Astronomy, University of Minnesota, 116
  Church Street, SE, Minneapolis, MN 55455; john@astro.umn.edu}
\altaffiltext{5}{School of Physics, University of Sydney, NSW 2006,
  Australia; agreen@physics.usyd.edu.au} 

\begin{abstract}

We present an analysis of rotation measure (RM) fluctuations from the
Test Region of the Southern Galactic Plane Survey (SGPS), along with
emission measure (EM) fluctuations in the same field taken from the
Southern H-Alpha Sky Survey Atlas. 
The structure function of RM fluctuations shows a relatively
steep slope at small scales (1~--~5 arcmin), a break in slope to a
flatter structure function at intermediate scales (5~--~60 arcmin),
and a systematic variation of the strength of fluctuations as
a function of position angle on the sky at the largest scales
(60~--~200 arcmin). The structure function of EM fluctuations shows
similar behavior, although the lower resolution of the data prevents
detection of a possible break in the spectrum. We interpret the
anisotropy in RM/EM structure on large scales as resulting from a
large-scale gradient in electron density (and possibly magnetic field)
across the region. The break in the slope of the RM structure
function at scales of $\sim5$~arcmin can be explained by contributions
from two spatially distinct magneto-ionized screens, most likely in
the Local and Carina spiral arms. The observed structure function then
implies that the outer scale of RM fluctuations in these screens is
$\sim2$~pc. Such behavior is in striking contrast to the expectation 
that interstellar turbulence forms an unbroken spectrum from kpc down
to AU scales.  We conclude that we have identified an additional source
of enhanced turbulence, injected on scales of a few pc, possibly seen
only in the Galactic plane.  The most likely source of such turbulence
is individual H\,{\sc ii} regions from relatively low-mass stars, whose
characteristic scale size is similar to the outer scale of turbulence
inferred here. These sources may be the dominant source of density and
velocity fluctuations in warm ionized gas in the Galactic plane.
\end{abstract}

\keywords{Galaxy: structure --- H~{\sc ii} regions --- ISM: structure
  --- techniques: polarimetric --- radio continuum: ISM --- turbulence}

\section{INTRODUCTION}

The evidence for the presence of turbulence in the interstellar medium
(ISM) is overwhelming. Although some studies of the characteristics of
this turbulence (viz.\ the inner and outer scales, shape of the
spectrum and power law spectral index) indicate the presence of
standard incompressible Kolmogorov (1941) turbulence, many
observations indicate different kinds of turbulence, drivers and
environments. Armstrong, Rickett \& Spangler (1995) compiled
observations of (among others) interstellar scattering of pulsars and
extragalactic sources, dispersion measures of pulsars 
and rotation measures (RM) of extragalactic sources in one power
spectrum. The result was the so-called ``big power law in the sky'', a
Kolmogorov-like power spectrum of electron density fluctuations over
12 orders of magnitude, from a fraction of an AU to kiloparsecs. 
Although the big power law in the sky is very well-determined
on small scales (up to $\sim 0.001$~pc), the extension towards larger
scales is based only on two types of measurements: the RM of extragalactic
sources, which gives an upper limit to the amount of structure in the
electron density because the contribution of the magnetic field is
unknown; and from velocity measurements of H\,{\sc i}, making
assumptions about the connection between neutral and ionized
material.

The dominant source for the large-scale energy input in the
Galaxy is generally assumed to be supernova explosions (Spitzer 1978;
Vollmer \& Beckert 2002; Korpi et al.\ 1999). Other possible
sources are superbubbles, massive \HII\ regions and massive stellar
winds \citep{nf96}, Galactic fountains, chimneys, or gravitational
scattering by transient spiral waves (see Sellwood \& Balbus (1999)
and references therein), gravitational instabilities in a shearing
disk (Elmegreen, Elmegreen \& Leitner 2003), or magneto-rotational
instabilities \citep{mk04}. The big power law in the sky certainly
suggests the classical scenario of turbulent energy input on large scales,
which cascades down to smaller scales until the energy is dissipated
on the smallest scale \citep{k41}.

However, there are indications of other types of turbulence in the
ISM as well. \cite{ms96} found a break in the structure function of
RMs of extragalactic sources and emission measures (EM) of the warm
ionized gas on scales of a few parsec, which they interpreted as a
transition from three-dimensional to two-dimensional turbulence as one
moves from small to large scales. Analytic theory (Goldreich \&
Sridhar 1995, 1997) and simulations (e.g.\ Cho, Lazarian \& Vishniac
2002; Maron \& Goldreich 2001) show 
that magnetic fields can lead to anisotropic turbulence, which is
predicted to exhibit the same power law spectral index as Kolmogorov
turbulence. Furthermore, observations of RMs of extragalactic sources
(Simonetti \& Cordes 1986; Spangler \& Reynolds 1990; Clegg \etal\
1992) show higher amplitudes of structure in RM in the Galactic plane
than out of the plane, suggesting the existence of an additional
source of structure on parsec scales.  Results from interstellar
scattering show a similar enhancement of structure in the inner
Galaxy, but on much smaller scales (Rao \& Ananthakrisnan 1984;
Dennison \etal\ 1984; Anantharamaiah \& Narayan 1988).

Structure in the ionized ISM can be studied by way of structure
functions (SFs). Earlier determinations of SFs of RM using
extragalactic sources or the diffuse synchrotron background yielded
results that varied widely with scale and area in the sky. Flat
structure functions (indicating no structure on the probed scales)
found by Simonetti, Cordes \& Spangler (1984) on scales larger than
$4\degr$ towards the Galactic pole were interpreted as showing
structure intrinsic to the extragalactic sources with a negligible
Galactic contribution. However, this cannot explain the flat SFs from
low-latitude extragalactic sources found by \citet{scs84} on scales
$\ga 4\degr$ and by \citet{ccs92} on scales $\ga 1\degr$.
Furthermore, Sun \& Han (2004) have recently found shallow SFs of RM
in the Galactic plane, and a flat SF at the North Galactic Pole, also
from extragalactic sources. Haverkorn, Katgert \& de Bruyn (2003a)
studied the structure function of RM from diffuse radio emission, and
found very shallow slopes for two fields at intermediate latitudes. 

In this paper we study the turbulent structure in ionized gas in
the inner Galactic plane, by means of SFs of RM from the diffuse
synchrotron background, and of SFs of EM from H$\alpha$ emission, both
in a region in the inner Galactic plane. 

In Section~\ref{s:obs} we present our radio polarimetric observations
of the Galactic synchrotron background in the Galactic plane. 
Section~\ref{s:sfdet} discusses the computation of the structure
function, while in Section~\ref{s:sfint} we interpret the structure
function as arising from Faraday screens in two spiral arms along the
line of sight, both exhibiting turbulent structure. In
Section~\ref{s:hii}, we discuss evidence for enhanced density
fluctuations in the Galactic plane, and propose that the structure is
dominated by discrete \HII\ regions.

%------------------------------------------------------------------
\section{THE SOUTHERN GALACTIC PLANE SURVEY TEST REGION}
\label{s:obs}

%***********************
\begin{figure*}
\centerline{\psfig{figure=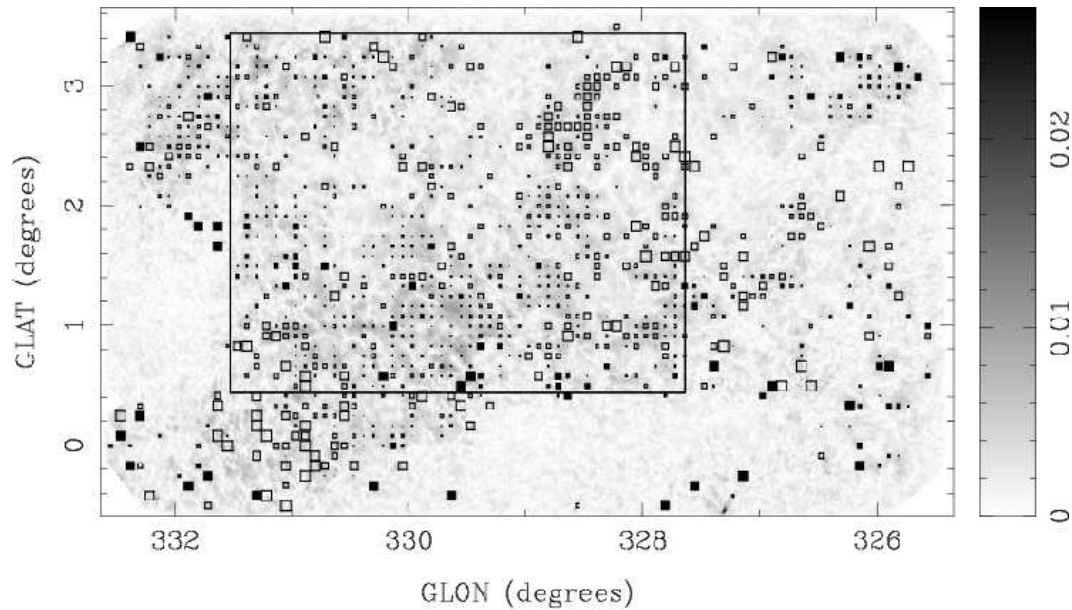,width=.85\textwidth}}
\caption{Rotation measure in the SGPS Test Region in squares, overlaid
  on polarized intensity in grey scale given in Jy/beam. Open squares
  denote negative RMs and filled squares positive ones. The length of
  a square is proportional to the magnitude of the RM for $|\mbox{RM}|
  < 100$~rad~m$^{-2}$, and constant for $|\mbox{RM}| \ge
  100$~rad~m$^{-2}$. RM values have been given only if S/N~$>5$ and
  reduced $\chi_r^2 < 2$. The rectangular box drawn in the figure
  shows the area over which structure functions are computed.}
\label{f:sgps}
\end{figure*}
%***********************
%***********************
\begin{figure*}
\centerline{\psfig{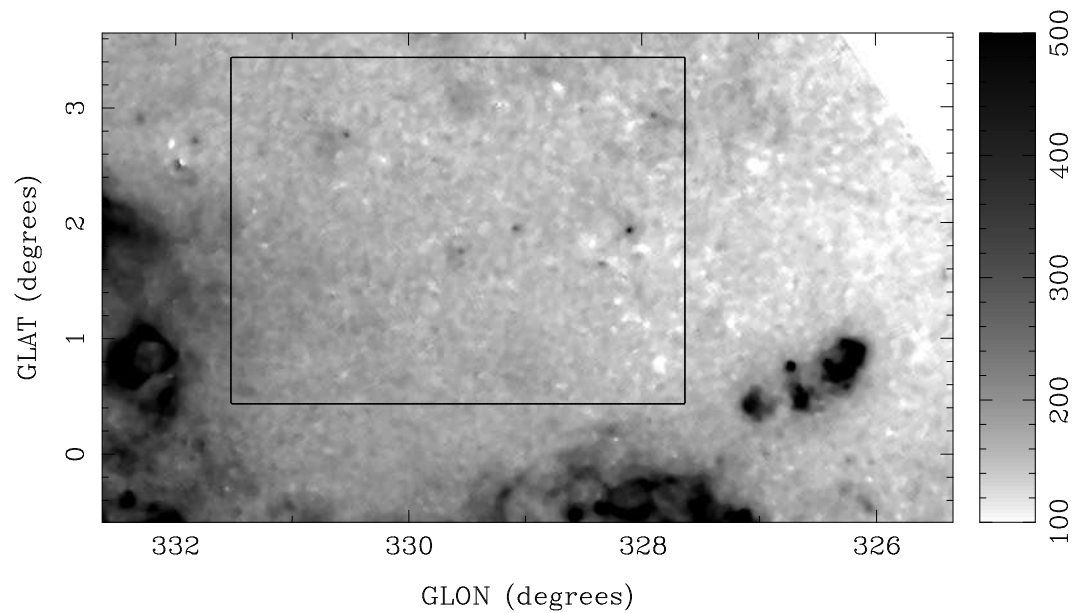}}
\caption{H$\alpha$ emission in the SGPS Test Region, from the Southern
  H-Alpha Sky Survey Atlas \citep{gmr01}. The scale is in
  deciRayleighs
  (1~Rayleigh~=~$10^6/4\pi$~photons~cm$^{-2}$~s$^{-1}$~sr$^{-1}$) and
  the drawn box is the same as in Figure~\ref{f:sgps}. A gradient
  in emission is visible from the bottom left corner to the top right
  corner of the box.}
\label{f:shassa}
\end{figure*}
%***********************
The Southern Galactic Plane Survey (SGPS) is a radio survey in H\,{\sc
i} and linearly polarized continuum at 1.4~GHz performed with the
Australia Telescope Compact Array (ATCA) and Parkes 64m dish
(McClure-Griffiths et al.\ 2001; Gaensler et al.\ 2001). The first
phase of the survey extends over $ 253^{\circ} < l <358^{\circ}$ and
$|b|<1.5^{\circ}$, has a spatial resolution of $1\arcmin$ and a
velocity resolution of 1~km~s$^{-1}$. Before embarking on the complete
survey, a Test Region was observed, spanning
$325.5^{\circ}<l<332.5^{\circ}$ and $-0.5^{\circ}<b<3.5^{\circ}$
\citep{gdm01}. The continuum data for the SGPS Test Region consists of
nine frequency bands each of 8~MHz in width, in the range 1336~MHz to
1424~MHz. From the linear polarization position angle $\phi$ at these
nine frequencies, RMs were derived as $\phi \propto
\mbox{RM}\lambda^2$, where RM~$= 0.81 \int n_e B_{\parallel} ds$ 
with $n_e$ the thermal electron density in cm$^{-3}$, $B_{\parallel}$
the component of the magnetic field parallel to the line of sight in
$\mu$G, and $ds$ the path length in parsecs. \citet{gdm01} give a
detailed analysis of the structure in linear polarization and RM in
the Test Region, focused on morphology and individual sources. RM
fluctuations were apparent in their Figure~7, but not analyzed. We use
their data, but use a slightly different algorithm to derive RMs.

Gaensler et al.\ use a standard algorithm within the data reduction
package
MIRIAD\footnote{http://www.atnf.csiro.au/computing/software/miriad}
\citep{sk03} 
to compute RMs, which they describe in detail. The MIRIAD algorithm
solves the problem of $n \pi$~radians ambiguity in the polarization
angle by first computing an RM value from the first two
frequencies. Then, the polarization angles of the other frequencies
are rotated $n \pi$~radians so that they lie closest to the angle
predicted from the first two frequencies, and the RM is recomputed
using these angles. Therefore, a large error in (one of) the two
first frequencies can yield the wrong determination of RM. In this
way, a small part ($\sim$~6\%) of the RM values Gaensler et al.\
computed was probably incorrect. Note that all these erratically
determined RMs had high reduced $\chi^2$ of the relation $\phi =
\mbox{RM}\lambda^2$, which caused them to be disregarded in their
analysis. Therefore the use of this algorithm did not lead to wrong
quoted values of RM, but rather to a small number of unjustly
disregarded RMs.

Here, we incorporate the ambiguity of polarization angle $\phi$ by
always taking the least possible angle difference between $\phi$ in
adjacent frequency bands, i.e.\ $|\phi_i - \phi_j| < \pi$~rad where
$i$ and $j$ are adjacent frequencies (Haverkorn, Katgert \& de Bruyn
2003b; Brown, Taylor \& Jackel 2003). The frequencies are so closely
spaced that this method 
renders the correct RM for all RM~$\la 1900 $~rad~m$^{-2}$. Since in our
field RMs are not higher than a few hundred rad~m$^{-2}$ and mostly
much lower, this is the case for all derived RMs. 

Figure~\ref{f:sgps} shows the RM in the Test Region, superimposed
on the continuum polarized intensity in grey scale. The resolution is
about 1$\arcmin$ and the data are about three times oversampled. Open
(filled) squares represent negative (positive) RM, and the size of a square
is proportional to the magnitude of RM for $|RM| < 100$~rad~m$^{-2}$
and constant for $|RM| \ge 100$~rad~m$^{-2}$. Only every third beam is
plotted for clarity. Furthermore, only ``reliably determined'' RMs are
shown, i.e.\ RMs for which the signal to noise ratio S/N~$>5$ and the
reduced $\chi^2$ of the linear $\phi(\lambda^2)$ fit was $\chi_r^2 <
2$. The dependence of $\phi$ on $\lambda^2$ can be non-linear due to
depolarization effects (Burn 1966; Sokoloff et al.\ 1998). Mild
non-linearity is allowed by including reduced $\chi_r^2$ up to
2. About 41\% of the data have low enough reduced $\chi_r^2$, of which
37\% has a high enough S/N. So about 15\% of the data has reliably
determined RMs.

As these data are taken solely with the ATCA interferometer, they lack
large-scale information ($\ga 30\arcmin$) in the observed Stokes $Q$
and $U$ maps. If a large component of $Q$ and $U$ is missing,
the computed values of polarization angle will be incorrect, and
therefore the RM will be in error as well. However, the polarization
angle depends very non-linearly on $Q$ and $U$, so that a large
missing component in $Q$ and/or $U$ would destroy the linear relation
between $\phi$ and $\lambda^2$. The fact that so many RM's with low
reduced $\chi^2$ are observed suggests that any missing large-scale
components should be small. This needs to be confirmed with data from
the single dish Parkes telescope, which will be the subject of future
studies.

We note that the inability to detect large-scale structure in Stokes
$Q$ and $U$ images due to the lack of single-dish data {\em does not}\
imply an inability to detect large-scale components in RM.  As long as
small-scale structure in RM exists superimposed on any large-scale RM,
the complete $Q$ and $U$ signals will be detected, which correspondingly
allow us to retrieve the {\em total}\ RM,  not just the small-scale
component. Therefore, fluctuations in RM can be probed at much larger
scales than the maximum spatial scale probed by the interferometer.

As RM only provides information on the product $n_eB_{\parallel}$
integrated over the path length, independent information on the path
length integrated distribution of thermal electrons is extremely
useful for estimation of the structure in the magnetic field. 
Therefore, we also used emission measure EM~$=\int n_e^2 ds$ data from
the Southern H-Alpha Sky Survey Atlas (SHASSA, Gaustad et al.\ 2001),
which provides comparable resolution H$\alpha$ data of the Southern
sky, see Figure~\ref{f:shassa}. Caution is needed in comparing RM and
EM, as the path lengths may not be the same, and the H$\alpha$ data is
not corrected for dust extinction. The SHASSA provides data which is
median filtered over four 0.8$\arcmin$ pixels to remove star
residuals, and has a sensitivity of about 0.5~Rayleigh (1~R~$ =
10^6/4\pi$~photons~cm$^{-2}$~s$^{-1}$~sr$^{-1}$, corresponding to
EM~$\approx 2$~cm$^{-6}$~pc at T~$=8000$~K).

%------------------------------------------------------------------
\section{DETERMINATION OF STRUCTURE FUNCTIONS}
\label{s:sfdet}

%***********************
\begin{figure*}
% from makesf1d.pro: sfrm_ind_bestfield_oplot.ps and ...
\centerline{\psfig{figure=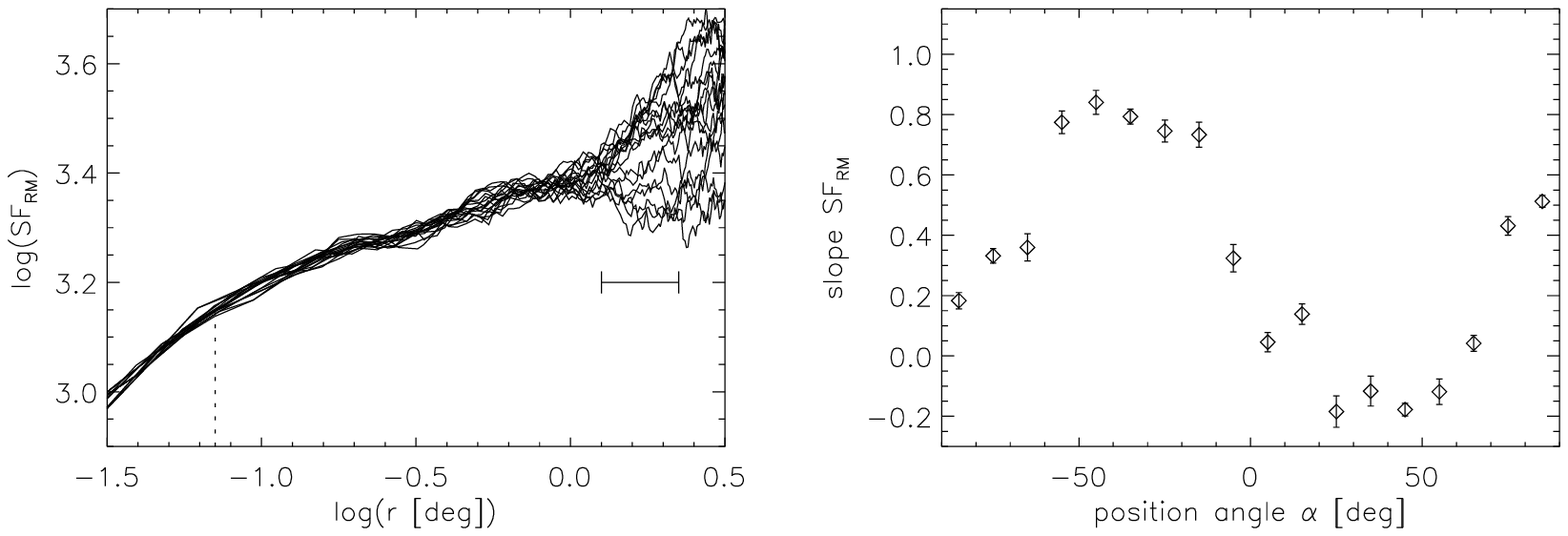,width=.9\textwidth}}
\centerline{\psfig{figure=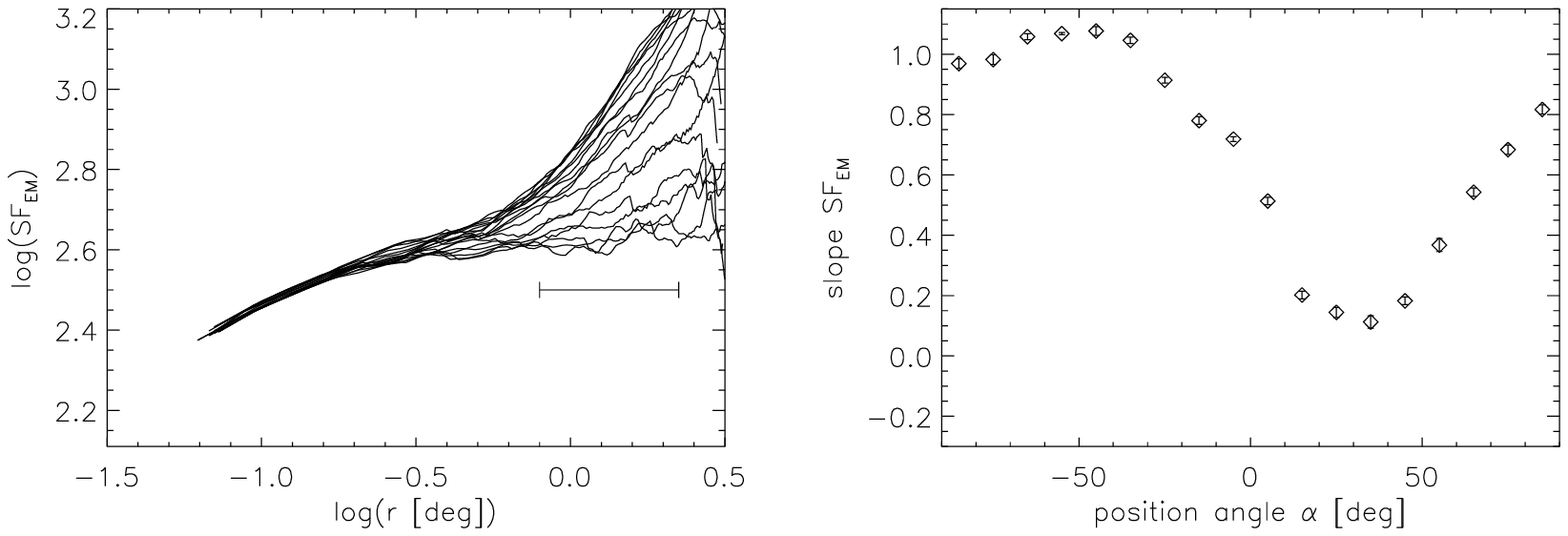,width=.9\textwidth}}
\caption{Top left: Structure functions of RM, \sfrm, in the SGPS Test
  Region for different position angles $0^{\circ}<\alpha<180^{\circ}$
  (positive $b$ through positive $l$) superimposed. The vertical
  dotted line shows the position of the break in slope, and the
  horizontal line gives the range over which the linear fit was
  performed to yield the slope given in the right hand plot. Top
  right: dependence of the slope of \sfrm\ on position angle
  $\alpha$. Bottom: The same for \sfem.} 
\label{f:sf}
\end{figure*}
%***********************

Ideally, the scales of variability in a signal such as RM are
determined by the power spectrum. However, as the RM data contain
blank spaces in the regular grid, we use structure functions (SFs) instead
(Simonetti et al.\ 1984; Minter \& Spangler 1996). We define the
second order structure function SF of RM and EM as a function of
distance lag $\mathbf{r}$ as  
\begin{eqnarray}
 \sfrm(\mathbf{r}) &=& \langle (\mbox{RM}(\mathbf{x}) -
 \mbox{RM}(\mathbf{x}+\mathbf{r}))^2 \rangle_{\mathbf{x}}  \\ 
 \sfem(\mathbf{r}) &=& \langle (\mbox{EM}(\mathbf{x}) -
 \mbox{EM}(\mathbf{x}+\mathbf{r}))^2 \rangle_{\mathbf{x}} 
\end{eqnarray}
where $\langle \rangle_{\mathbf{x}}$ denotes the averaging over all
positions $\mathbf{x}$ in the field. 

Structure functions of RM and EM were determined in the area of the
SGPS Test Region for which the most data were available, i.e.\ within the
box drawn in Figs.~\ref{f:sgps} and~\ref{f:shassa}. The noise was
taken into account by subtracting a SF of the error $\delta$ 
in the RM data from \sfrm, as explained in the Appendix. We
computed one-dimensional SFs for different position angles $\alpha$
(positive $b$ through positive $l$) of the distance lag vector
$\mathbf{r}$, with an interval of $\Delta\alpha=10^{\circ}$.

The left hand plots in Figure~\ref{f:sf} give the SFs of RM (top) and EM
(bottom) for different position angles $0^{\circ}<\alpha<180^{\circ}$
superimposed for regular sampling in $r$. Errors in the SF are
typically $\log(\sfrm)\la 0.05$. The two SFs agree remarkably well:
both are very shallow; the slope of \sfrm\ is $\sim0.2$ and the slope
of \sfem\ $\sim0.25$ for $4^{\prime}\la r\la5^{\circ}$. The slope of
\sfrm\ shows a break, below which the SF steepens to $\sim0.5$.
Furthermore, at larger scales an anisotropy occurs with position angle
in both SFs. A linear fit to the slope at large scales was performed
for each position angle over the range given by the horizontal lines
in the left hand plots. The slopes of these linear fits are given in
the right hand plots of Figure~\ref{f:sf}. The anisotropy on large
scales in the SFs of both RM and EM is more or less sinusoidal, with a
maximum for RM and EM at the same position angle $\alpha\approx
-50^{\circ}$.

The sudden change in slope on scales smaller than the break is
probably not due to resolution effects in the RM data, as the
resolution of the RM data is FWHM~$= 86\arcsec$ ($\log(r) \approx
-1.6$), which is well below the break. In the EM data, a median
filtering over 4 pixels means that at scales $\log(r) > -1.17$,
beams are independent, so that the position of the break in
\sfrm\ falls below the resolution of the H$\alpha$ data. The 
unsmoothed data (with a resolution of 0.8\arcmin) are too contaminated
with point sources to determine if a break in the spectrum is present
in those data.

The anisotropy in the SF slope at large scales is not an artifact of
the gridding or the coordinate system, as its maximum does not coincide
with one of the axes. Furthermore, we constructed one-dimensional SF's
of the same data in (RA,~dec) coordinates. The slopes of these SF's show
maxima at the same intrinsic position angle as the data in Galactic
coordinates, confirming that the anisotropy is not an artifact of the
gridding of the data. The same analysis has also been applied to maps
of the Stokes $Q$ and $U$ parameters, and to a map of randomly
scrambled RM values in which the observed RMs are redistributed at
random positions in the field. The SF's of none of these maps showed
an anisotropy, again indicating that the change in slope of SF with
direction is physical and not an artifact.

%------------------------------------------------------------------
\section{INTERPRETATION OF THE STRUCTURE FUNCTIONS}
\label{s:sfint}

\subsection{The Anisotropy of the SF Slope at Large Scales}

Two very different characteristics of a medium can produce an
anisotropy of the slopes of SFs: anisotropic turbulence and a
large-scale gradient in the medium.

If a strong enough large-scale magnetic field is present in the
turbulent ionized medium, the turbulence will be anisotropic (e.g.\
Goldreich \& Sridhar 1995, 1997; Cho \& Vishniac 2000). Observing
anisotropic turbulence through RM data is believed to be difficult or
even impossible because the anisotropy is directed with respect to the
{\it local} magnetic field \citep{clv02}, so that structure in the
magnetic field along the line of sight is likely to destroy any
observational effects of the anisotropic turbulence. If anisotropic
turbulence is present in the SFs, it should be seen on all scales
instead of only large ones. Furthermore, anisotropic turbulence would
yield a set of SFs as a function of position angle which have the same
amplitude of fluctuations but on different scales. This is not the
case in our observations, where slopes vary from $\sim$~1 to below
zero with position angle. Therefore, we do not believe that the
anisotropy in the slope of the SF is caused by anisotropic turbulence. 

If the anisotropy in the slope is caused by a gradient across the
field, one would expect a steep SF slope in the direction of the
gradient, and a flat SF slope in the perpendicular direction. Noise
and the superposition of a regular structure like a 
sinusoid can decrease the slope to below zero. This variation in slope
to values below zero is exactly what we observe in both \sfrm\
and \sfem. In fact, in the H$\alpha$ data the gradient can be directly
seen in the maps (Figure~\ref{f:shassa}). An H$\alpha$ map of a larger
region shows a complex of \HII\ regions and supernova remnants
extending over more than 15 degrees in Galactic longitude at higher
longitude and lower latitude from the SGPS Test Region, and the
EM from that complex decreasing towards and across the Test
Region. This coincides with the direction of the gradient computed
from the SFs.

%***********************
\begin{figure}
% from scatter_rmem.pro
\centerline{\psfig{figure=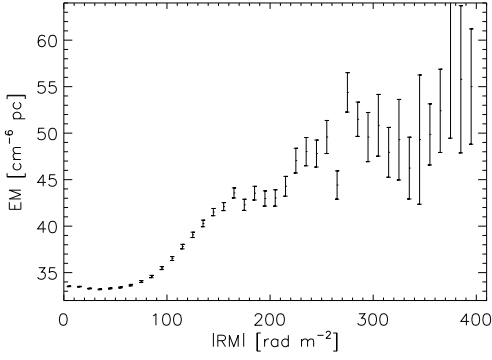,width=.45\textwidth}}
\caption{Average EM in bins of absolute RM with width
  $\Delta |$RM$|$~=~10~rad~m$^{-2}$. The errors are the errors in the
  mean within a bin.} 
\label{f:rmem}
\end{figure}
%***********************
A maximum amount of fluctuations in RM at the position angle
$-50^{\circ}$ does not determine if the gradient is directed towards
$-50^{\circ}$ or $130^{\circ}$. To find the direction of the RM
gradient, we looked at the correlation between the magnitude of EM and
RM. Figure~\ref{f:rmem} shows the average EM in bins of $|$RM$|$ with 
width $\Delta$RM~=~10~rad~m$^{-2}$. The magnitudes of EM and RM are
clearly correlated, indicating that the gradient in $|$RM$|$ follows the
gradient in EM. Therefore $|$RM$|$, like EM, decreases towards
$\alpha=-50^{\circ}$. This is to be expected if gradients in electron
density and/or path length dominate, and the magnetic field either
shows the same gradient or does not change significantly over the
field.

%***********************
\begin{figure}
% from sf/makesf1d_modelanddata.pro
\centerline{\psfig{figure=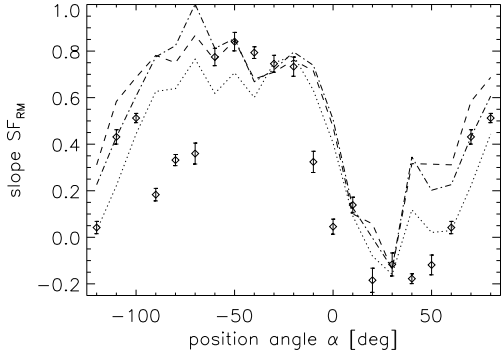,width=.45\textwidth}}
\caption{Slope of \sfrm\ as a function of position angle $\alpha$
  (diamonds), and models of a linear gradient (dotted), a broad
  sinusoid (dashed), and a Gaussian (dash-dotted). A noise
  component of 5~rad~m$^{-2}$ has been added to the model RMs,
  comparable to the noise in the data.}
\label{f:models}
\end{figure}
%***********************
To determine what kind of gradient could produce an anisotropy in the
slope of SF as observed, we modeled the anisotropy in \sfrm\ 
using three simple functions: a linear gradient, a broad sinusoid and
a Gaussian decrease, all with the gradient oriented at
$\alpha=-50^{\circ}$. Figure~\ref{f:models} shows the results of the
modeling with each of these three functions, compared to the data
given by the diamonds. The linear gradient is given by the dotted
line, the sinusoid by the dashed line, and the Gaussian is denoted by
the dash-dotted curve. Noise is included in the models at
RM$_{noise}=5$~rad~m$^{-2}$, similar to the noise level in the data.
These simplified models show an anisotropy in the slope of comparable
amplitude and minimum to the data, although we cannot distinguish
between them. More elaborate modeling of the anisotropy in the SF
slope is beyond the scope of this paper.

The three models fit best for a total gradient in RM across the field
of $\Delta RM\approx 3$~rad~m$^{-2}$, while the H$\alpha$ data show a
change in EM of about  10~cm$^{-6}$~pc ($\sim$~5~Rayleigh at $T\approx
8000$~K). These changes can be due to gradients in electron density or
the path length, and most likely a combination of these, possibly
accompanied by a gradient in magnetic field. Therefore the measured
$\Delta RM$ and $\Delta EM$ give too few constraints to estimate the
gradients separately in magnetic field strength and electron density.

\subsection{The Shallow Slope and Break}

%***********************
\begin{figure}
\centerline{\psfig{figure=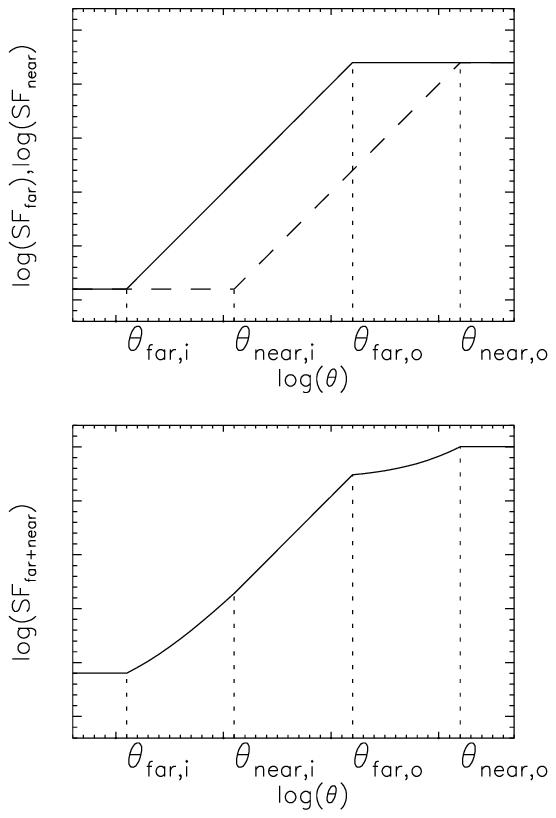,width=0.45\textwidth}}
\caption{Example of the superposition of two Kolmogorov-like structure
functions. Top: two structure functions of turbulent Faraday screens
with an outer scale $r_o$ and an inner scale $r_i$, at distances
``near'' (dashed) and ``far'' (solid). Bottom: the addition of the two
structure functions in the top panel shows a break in the slope, with
a shallower slope at large scales.}
\label{f:ex}
\end{figure}
%***********************

A structure function of RM produced by a slab of Kolmogorov turbulence
with an inner scale $r_i$ and an outer scale $r_o$ is shown as the
solid line in the upper panel of Figure~\ref{f:ex} (e.g. Stinebring \&
Condon 1990). The SF is constant on small scales due to dissipation at
the inner scale, and then starts to rise at $r_i$. If the spatial
fluctuation in RM is a statistically homogeneous and isotropic random
variable and the spectrum of RM is a power law with spectral index
$s$, the slope of the SF $b$ ($\sfrm \propto r^b$ for $r_i < r < r_o$) is
related to $s$ as \citep{scs84}:
\begin{equation}
  \sfrm(r) \propto r^b \hspace*{0.2cm} \mbox{where} \;
  b        = \left\{ \begin{array}{ll}
                  s - 2 & \;\;2 < s < 4 \\
                  2          & \;\;s > 4 \\
                \end{array}
                \right.
\end{equation}
Therefore a Kolmogorov spectrum ($s=11/3$) should exhibit a structure
function slope $b = 5/3$. Above the outer scale of structure $r_o$,
the structure function saturates to the constant value of
$2\sigma^2_{\mbox{\sc rm}}$, where $\sigma^2_{\mbox{\sc rm}}$ is the
variance of the RM distribution.

However, the observed SFs do not show this behavior. In the observed
\sfrm, a break is visible, and the slope at scales larger than the break
is very shallow. This behavior can be most easily explained by the
superposition of two Faraday screens at different distances, both with
a Kolmogorov-like SF and with certain physical inner and outer
scales. Even if the {\em physical} scales of the structure in these
screens are equal, the RM will exhibit structure on different {\em
  angular} scales, as shown in Figure~\ref{f:ex}. The upper panel
shows two Kolmogorov-like SFs with the same amplitude at different
angular scales $\theta$. In the lower panel the sum of the two SFs is
shown, revealing both a break and a shallow spectrum, as observed in
Figure~\ref{f:sf}.

Due to a larger electron density and magnetic field strength in the
spiral arms (Beck et al.\ 1996; Frick et al.\ 2001; Cordes \& Lazio
2003), the major contribution to the RM occurs in the spiral
arms. \citet{gdm01} determined that the observed polarized emission
originates predominantly from the Crux arm at a distance of 3.5~kpc
from the Sun. The major contribution to the RM is most likely made
by the Carina and Local spiral arms, located at 100~pc and 1.5~kpc
from the Sun, respectively. Then, the Local arm corresponds to the
nearby screen in Figure~\ref{f:ex}, and the Carina arm to the far
screen. The break in the SF can be interpreted as the
location of the outer scale of structure in the far screen, and
therefore denotes the outer scale of the fluctuations in the Carina
spiral arm with  $r_{\mbox{\it\scriptsize far,o}}\approx
2$~pc. Similarly, a lower limit to the outer scale of fluctuations in
the nearby screen is given by the position at which the large-scale
anisotropy starts dominating the spectrum, i.e.\ $\log(r) = 0$. This
corresponds to the outer scale of structure in the Local Arm
$r_{near,o} \ga 1.7$~pc. The shallow slope of \sfem\ confirms
this hypothesis, as the H$\alpha$ emission in this direction is
believed not to originate farther away than 1~--~2~kpc, so all
H$\alpha$ emission from the Crux arm would be absorbed by intervening dust.
Note that \citet{ev03} demonstrated that the autocorrelation length of
the Galactic magnetic field is smaller than the autocorrelation length
of the RM in typical astrophysical situations, so that the given
values for the outer scale of structure could be an upper limit.

%------------------------------------------------------------------
\section{\HII\ REGIONS AS THE DOMINANT SOURCE OF STRUCTURE}
\label{s:hii}

The only other comparable study of EM and RM fluctuations is that made
by \citet{ms96}. Minter \& Spangler found an outer scale for
three-dimensional turbulence in ionized gas of around 4~pc in a region
of about 100 square degrees centered at $(l,b) = (143\degr,
-21\degr)$, but concluded that two-dimensional turbulence extended to
larger scales of at least several hundred parsecs. Similarly, \citet{sc86}
found structure in the RM of extragalactic point sources in a region
at $70\degr < l < 110\degr$, $-45\degr < b < 5\degr$ on scales of
$4\degr$ up to $40\degr$, equivalent to 35 to 400~pc at a distance of
2~kpc. Studies of velocity fluctuations in neutral gas also provide
evidence for turbulence out to scales of 100~--~1000~pc (Larson 1979;
Padoan et al.\ 2001).

These results are all in striking contrast to the structure function
analyses presented in Section~\ref{s:sfdet}, where we provide
evidence that the outer scale for fluctuations in the density of
ionized gas is only $r_o = 1-2$~pc.  Because of the ubiquity of the
turbulent cascade otherwise seen in ionized gas over at least
12~orders of magnitude in scale \citep{ars95}, we are reluctant to
interpret our result as indicating that a different physical mechanism
is operating in our particular region. Rather, we interpret our
results as being due to an additional component of turbulence in the
warm ISM. We propose that there are two contributions to turbulence in
ionized gas: the overall cascade from the very largest Galactic scales
described by \citet{ars95} found throughout the ISM; and a more
localized source of turbulence with an outer scale of $\sim$~2~pc,
possibly of much stronger amplitude, found only in the Galactic plane.

There is already good evidence to support such a possibility.  The
previous analyses of fluctuations in H$\alpha$ emission and in Faraday
rotation which have argued for an outer scale $r_o \ga 100-1000$~pc,
have focused on sources at high Galactic latitudes, in the outer Galaxy,
or in specific localized regions. However, studies of turbulence
including the ionized gas in the Galactic plane of the inner Galaxy
generally show enhanced turbulence in those regions.
\citet{ccs92} considered RMs of extragalactic sources in the inner
Galaxy and at low latitude ($45^\circ < l < 93^\circ$, $|b| 
< 5^\circ$).  Because their background sources provided only sparse
and irregular sampling, Clegg et al.\ were unable to calculate
a continuous structure function with a well-defined slope, as we have
done here. However, their data provide clear evidence that at scales
$\ga1^\circ$, the amplitude of RM fluctuations at low latitudes is a
factor $>10^3$ stronger than those seen at high latitudes. 
Clegg et al.\ concluded that this results from enhanced turbulence at
low latitudes, generated by discrete structures in the Galactic
plane. \citet{sc86} also argued for an additional contribution to the
turbulence in the Galactic plane by comparing RMs of extragalactic
sources in and out of the Galactic plane. 

A similar conclusion was reached by \citet{sr90}, who compared the EMs
towards eight extragalactic sources with the sizes of these sources'
scattering disks as measured by VLBI. They found that heavily
scattered sources were the same sources which also showed 
enhanced EMs along their sight lines. They concluded that heavily
scattered sources are viewed through an additional component of
ionized gas, with turbulent properties distinct from those of the
diffuse ionized medium through which weakly scattered sources are
seen. Spangler \& Reynolds suggested that the additional component is
individual \HII\ regions in the Galactic plane. Furthermore, studies
of interstellar scintillation of pulsar signals and angular broadening
of extragalactic radio sources indicate higher electron density
fluctuations in the inner Galaxy than in the solar neighborhood (Rao
\& Ananthakrishnan 1984; Dennison et al.\ 1984; Cordes, Weisberg \&
Boriakoff 1985).

Several studies have tried to characterize the nature of this
additional component of structure in the inner Galaxy. \citet{cwb85}
concluded from pulsar scattering measurements that the ionized plasma
had both clumped and nearly uniform components. They suggested that the
enhanced scattering in the clumped component could be due to
ionization fronts associated with \HII\ regions or shocks associated
with stellar winds and supernova shells. Three different components
have been identified by \citet{ps99}, based on observations of pulsar
scintillation timescales, pulse broadening and dispersion measures. 
They distinguish a
component~A as the diffuse gas at high latitudes, which is
statistically uniform; a clumpy component~BI which is distributed
uniformly in the spiral arms and may correspond to Str\"omgren spheres
of O7 to B0 stars; and a clumpy component BII which is concentrated in
``compact regions'' ($\sim1.5$~kpc) and is associated with known \HII\
regions or supernova remnants. A similar subdivision was made by
\citet{eb93} for the external spiral galaxy NGC~6946. Ehle \& Beck created a
model explaining Faraday  rotation, depolarization and thermal radio
emission as observed at various wavelengths from $\lambda$2.8~cm to
$\lambda$20.5~cm. Their best fit model contains ionized gas in three
components: (1) the diffuse ionized gas with low density and high
filling factor; (2) the classical giant \HII\ regions with high
density and low filling factor; and (3) small ($\sim 1$~pc)
low-density \HII\ regions with intermediate filling factor.

Our measurements appear to confirm this earlier work, in that we similarly
have identified a region for which the characteristics of electron
density fluctuations are distinct from those found at high latitudes
and in quiescent regions. While \citet{sc86}, \citet{sr90} and
\citet{ccs92} showed that the {\em amplitude}\ of density fluctuations
was enhanced on parsec scales in the inner Galaxy, we have
demonstrated that the {\em outer scale}\ of such fluctuations is also
widely different. We note that Spangler \& Reynolds suggested that
the parameter which differentiates the two types of turbulence is
$\alpha^3/r_o$, where $\alpha$ is the standard deviation of the
density fluctuations divided by the rms density, and represents the
depth of density 
modulation due to turbulence. Spangler \& Reynolds estimated $r_o
\approx 180~\alpha^3$~pc for diffuse ionized gas, and $r_o \approx
1.0~\alpha^3$~pc for regions of enhanced scattering. Assuming strong
turbulence ($\alpha\approx 1$) for the regions of strong scattering,
they found a typical scale of 1~pc, which is very similar to what we
observe here. Other indications that  an outer scale of structure of a
few parsec may exist in the Galactic plane are presented by
\citet{hkb03a}. They show SFs of RM from the diffuse synchrotron
background in two regions in the second quadrant at latitudes of
$8\degr$ and $16\degr$. Because they determined the RM from
observations at 350~MHz, only the nearby medium ($\sim 500$~pc) is
probed, i.e.\ only the thin disk. The SFs computed by Haverkorn \etal\
may exhibit a break at scales of about 2~pc, and are flat on larger
scales. Furthermore, the \sfrm\ of extragalactic sources presented
by Lazio, Spangler \& Cordes (1990) seem to saturate on scales of a
few parsec, although 
they do not mention this feature. Finally, \citet{sc86} discuss RMs
from a region in the Galactic plane and show that there is no
structure in this region on scales above $4\degr$.

We therefore confirm the supposition made by earlier authors that in the
denser and more complicated regions of our Galaxy found at low latitudes,
the dominant source of injection for turbulence appears to be relatively
small, discrete, sources.  The density fluctuations induced by these
sources dominate the turbulent cascade generated at much larger scales
which is seen at higher latitudes.

%***********************
\begin{figure*}
{\psfig{figure=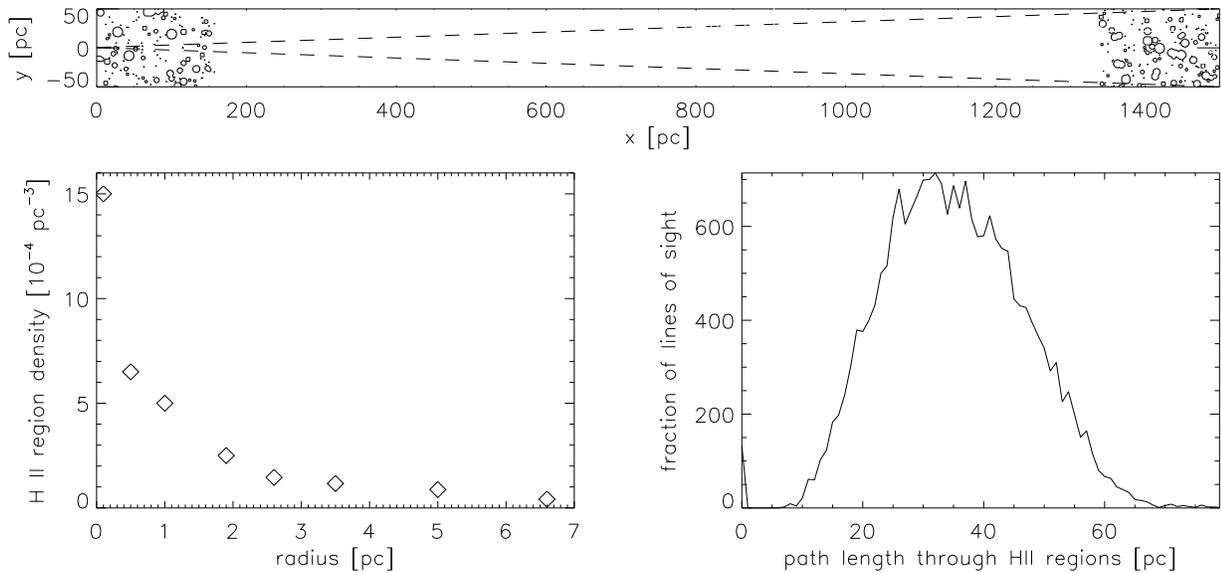,width=\textwidth}}
\caption{Top: two-dimensional cut through a volume distribution of \HII\
regions of B3 to A0 stars. The observer is located at $(x,y)=(0,0)$,
and the dashed lines indicate the field of view of the part of 
the SGPS Test Region used for the structure function analysis. Stars
are randomly positioned in the Local and the Carina spiral
arms. The circles indicate the sizes of the individual Str\"omgren
spheres. Bottom: number of \HII\ regions per cubic parsec as a
function of their radii (left) and the amount of the line of sight
taken up by \HII\ regions for many lines of sight in the field of view
(right).}
\label{f:hii}
\end{figure*}
%***********************

What are the sources likely to be injecting such turbulence?  Ionized gas
in the Galactic plane is dominated by the contributions from individual
\HII\ regions from massive stars.  While such sources span a very wide
range of sizes, depending on the mass and number of the central powering
stars and on the ambient density, there is reasonable evidence that
the characteristic scale for such sources is 1~--~2~pc, as implied here.
The radius of an idealized Str\"omgren sphere for the late-type
B~stars which most likely dominate the population of photo-ionizing
sources in the Galaxy is indeed a few pc (e.g.\ Prentice \& ter Haar
1969), while measurements of the sizes of extragalactic \HII\ regions
show that the distribution indeed peaks at diameters $<5-10$~pc (Hodge,
Lee \& Kennicutt 1989; Paladini, Davies \& DeZotti 2003). 

If the observed structure is mainly due to \HII\ regions from B~stars,
their number should be sufficient to cover the entire field of
view. We can estimate the number of B3 to A0~stars per square parsec
of the disk from the Present Day Mass Function (PDMF) presented by
\citet{ms79}, and the size of their \HII\ regions from \citet{pt69}
(who assume a mean density $n = 1~$cm$^{-3}$).
Stars more massive than B3 are excluded because of their scarceness,
and stars later than A0 because their \HII\ regions are of negligible
size. We construct a three-dimensional model of our field of view,
placing B3 to A0 stars randomly in the Local and Carina spiral arms
according to the PDMF, and no stars anywhere else. We assume a
thickness of 160~pc for both arms (twice the scale height) and
the Sun positioned at the far side of the Local arm. The resulting
distribution of \HII\ regions is shown in Figure~\ref{f:hii}. The upper
plot shows a two-dimensional cut through the model volume, where the
$x$-axis is directed along the line of sight, the
observer is located at $(x,y) = (0,0)$ and the dashed lines give the
field of view of the part of the SGPS Test Region used for the
structure function calculation. The lower left plot shows the
distribution of the sizes of \HII\ regions, and the lower right plot
presents how much of the line of sight is occupied by \HII\ regions,
for all independent beams in the region. Clearly, the number of \HII\
regions is enough to dominate the amount of structure at every line of
sight.

Interestingly, we note that \citet{ars95} calculated the contribution
of the warm ionized ISM to interstellar turbulence, and concluded that
spheres of density $n_e \sim 0.2$~cm$^{-3}$ and radius $\sim2$~pc
could produce fluctuations comparable to those observed. Armstrong et
al.\ suggested that the amplitudes of fluctuations on such scales were
consistent with the turbulent cascade seen at both larger and smaller
scales.  A calculation of whether our data contain evidence for higher
amplitude than or comparable fluctuations to those found elsewhere
will be the subject of a later paper in this series.

\citet{mk04} argue that the amount of energy input from \HII\ regions
is only a few percent of the energy needed to drive the turbulence in
diffuse ionized gas. However, the additional component of turbulence that
we observed could be almost entirely due to structure within the \HII\
regions themselves, with a negligible contribution from the turbulence
in the diffuse warm ISM (Spangler \& Reynolds 1990; Clegg et al.\ 1992). 

If individual \HII\ regions dominate the structure, what is the
expected SF spectral index $b$? The additional structure component in
the ionized gas could be due to discrete edges of \HII\ regions,
turbulence inside the \HII\ regions, turbulence in the ISM invoked by
the \HII\ regions, or a combination of these. For structure in RM
caused only by discrete edges of \HII\ regions \sfrm\ is expected to
exhibit a slope $b=2$ (Rickett, private communication). There is ample
observational evidence that \HII\ regions are turbulent (O'Dell 1991;
Joncas 1999; Gaensler et al.\ 2001), but this turbulence may operate
on much smaller scales of $\sim 0.1$~pc (Joncas 1999). The observed
steepening of the  \sfrm\ slope below $\log(r)\approx-1.1$ possibly
constitutes a transition region to steeper slopes of $b = 2$, or to
Kolmogorov-like turbulence ($b=5/3$) on scales $\la 2$~pc.

%------------------------------------------------------------------
\section{CONCLUSIONS}
\label{s:con}

We have analyzed structure functions (SF) of RM and EM data in the
SGPS Test Region. These show very consistent results: 
\begin{itemize}
\item the SFs of both RM and EM, \sfrm\ and \sfem, exhibit a linear
  slope in log-log space, with \sfrm\ showing a break at
  $\sim4\arcmin$ to a flatter slope at larger angular scales; 
\item the slope at angular scales $\ga4\arcmin$ is shallow though
  non-zero in both \sfrm\ and \sfem, with a SF spectral index $b
  \approx 0.2$; 
\item at scales larger than about a degree, the spectral index of
  \sfrm\ and \sfem\ is anisotropic, forming a quasi-sinusoidal
  dependence on position angle. 
\end{itemize}
The anisotropy in one-dimensional SFs at large scales is explained by
a large-scale gradient in electron density, possibly accompanied by
magnetic field, across the field of view due to a foreground
structure. The break in \sfrm\ and shallowness of the slope
in both \sfrm\ and \sfem\ at smaller scales can be explained by
the superposition of two contributions to EM and RM of the Local and
Carina spiral arms. Within such a model, we infer the 
outer scale of structure in the spiral arms to be about 2~pc. As
Kolmogorov-like turbulence is observed in the ISM on scales much
larger than a few pc, these results constitute evidence for an
additional contribution to turbulent fluctuations in the Galactic
plane. The inferred outer scale of fluctuations agrees with the size
of Str\"omgren spheres around the most abundant late-type
B~stars. Therefore, \HII\ regions could provide the dominant source of
structure on pc scales in the Galactic spiral arms.

The suggestion that the break in \sfrm\ and the shallowness of
both \sfrm\ and \sfem\ is due to two
separate Faraday screens will be tested in a forthcoming paper by
studying SFs in different regions of the sky. This can also shed light
on the occurrence of any spatial variations of the turbulent
spectrum. Furthermore, the combined RM and EM data should enable at
least partial decoupling of magnetic field and electron density, and
the amplitudes of the SF will yield information on this additional
component of structure in the ionized ISM in the Galactic plane.

\acknowledgments
We thank Ellen Zweibel, Alex Lazarian, Jungyeon Cho, Dipanjan Mitra
and Chris Brunt for stimulating discussions and helpful suggestions.
The Australia Telescope is funded by the Commonwealth of Australia for
operation as a National Facility managed by CSIRO. The Southern
H-Alpha Sky Survey Atlas (SHASSA) is supported by the National Science
Foundation. MH and BMG acknowledge the support of the National Science
Foundation through grant AST-0307358. 

\appendix
\section{NOISE CONTRIBUTION TO THE STRUCTURE FUNCTION}
For a rotation measure RM~$\pm~\delta$~rad~m$^{-2}$, the SFs of the RM
and the error $\delta$ can be separated using
\begin{eqnarray}
\mbox{SF}(\mbox{RM}_{obs})\!\!\!\!&=&\!\!\!\! \left< [(\mbox{RM}(x)+\delta(x)) 
                 - (\mbox{RM}(x+r)+\delta(x+r))]^2 \right> \nonumber \\
	     &=& \left< \mbox{RM}^2(x)+\mbox{RM}^2(x+r)
                 - 2\mbox{RM}(x)\mbox{RM}(x+r)\right>+\nonumber\\
             & & \left< \delta^2(x)+\delta^2(x+r)- \right.
		 \left.2\delta(x)\delta(x+r)\right> \nonumber \\
             &=& \mbox{SF(RM)} + 2\left<\delta^2\right> \\
	     &=& \mbox{SF(RM)} + \mbox{SF}(\delta)
\end{eqnarray}
since $\left<\mbox{RM}\delta\right> = 0$ and
$\left<\delta(x)\delta(x+r)\right>=0$. Therefore, the contribution of noise
in RM (which is assumed Gaussian) can be taken into account by
subtracting a SF of a Gaussian distribution of noise with width
$\sqrt{\left<\delta^2\right>}$ from the SF of the observed RM.


\begin{thebibliography}{}
\bibitem[Anantharamaiah \& Narayan(1988)]{an88} Anantharamaiah,
  K. R., \& Narayan, R. 1988, in ``Radio wave scattering in the
  interstellar medium'', ed.\  J.~M. Cordes, New York, American
  Institute of Physics, p. 185. 
\bibitem[Armstrong et al.(1995)]{ars95} Armstrong, J. W., Rickett,
  B. J., \& Spangler, S. R. 1995, \apj, 443, 209
\bibitem[Beck et al.(1996)]{bbm96} Beck, R., Brandenburg, A., Moss,
  D., Shukurov, A., \& Sokoloff, D. 1996, ARA\&A, 34, 155
\bibitem[Brown et al.(2003)]{btj03} Brown, J. C., Taylor, A.~R., \&
  Jackel, B. J. 2003, \apj, 145, 213
\bibitem[Burn(1966)]{b66} Burn, B. J. 1966, \mnras, 133, 67
\bibitem[Cho et al.(2002)]{clv02} Cho, J., Lazarian, A., \& Vishniac,
  E.~T. 2002, \apj, 564, 291
\bibitem[Cho \& Vishniac(2000)]{cv00} Cho, J., \& Vishniac,
  E.~T. 2000, \apj, 539, 273
\bibitem[Clegg et al.(1992)]{ccs92} Clegg, A. W., Cordes, J. M.,
  Simonetti, J. M., \& Kulkarni, S. R. 1992, \apj, 386, 143
\bibitem[Cordes \& Lazio(2003)]{cl03} Cordes, J.~M., \& Lazio,
  T.~J.~W. 2003, preprint (astro-ph/0301598) 
\bibitem[Cordes et al.(1985)]{cwb85} Cordes, J.~M., Weisberg, J.~M.,
  \& Boriakoff, V. 1985, \apj, 288, 221
\bibitem[Dennison et al.(1984)]{dbt84} Dennison, B., Broderick, J.~J.,
  Thomas, M., Booth, R.~S., Brown, R.~L., \& Condon, J.~J. 1984,
  \aap, 135, 199
\bibitem[Ehle \& Beck(1993)]{eb93} Ehle, M., \& Beck, R. 1993, \aap,
  273, 45
\bibitem[Elmegreen et al.(2003)]{eel03} Elmegreen, B.~G., Elmegreen,
  D.~M., \& Leitner, S.~N. 2003, \apj, 590, 271
\bibitem[En\ss lin \& Vogt(2003)]{ev03} En\ss lin, T.~A., \& Vogt,
  C. 2003, \aap, 401, 835
\bibitem[Frick et al.(2001)]{fss01} Frick, P., Stepanov, R., Shukurov,
  A., \& Sokoloff, D. 2001, \mnras, 325, 649
\bibitem[Gaensler et al.(2001)]{gdm01} Gaensler, B.~M., Dickey, J.~M.,
  McClure-Griffiths, N.~M., Green, A.~J., Wieringa, M.~H., Haynes,
  R.~F. 2001, \apj, 549, 959
\bibitem[Gaustad et al.(2001)]{gmr01} Gaustad, J.~E., McCullough,
  P.~R., Rosing, W., \& Van Buren, D. 2001, \pasp, 113, 1326
\bibitem[Goldreich \& Sridhar(1997)]{gs97} Goldreich, P., \&
  Sridhar, S. 1997, \apj, 485, 680
\bibitem[Goldreich \& Sridhar(1995)]{gs95} Goldreich, P., \&
  Sridhar, S. 1995, \apj, 438, 763 
\bibitem[Haverkorn et al.(2003a)]{hkb03a} Haverkorn, M., Katgert, P.,
  de Bruyn, A. G. 2003a, \aap, 403, 1045 %SFs
\bibitem[Haverkorn et al.(2003b)]{hkb03b} Haverkorn, M., Katgert, P.,
  de Bruyn, A. G. 2003b, \aap, 403, 1031 % auriga
\bibitem[Hodge et al.(1989)]{hlc89} Hodge, P., Lee, M.~G., \&
  Kennicutt, R~C., Jr. 1989, \pasp, 101, 32
\bibitem[Joncas(1999)]{j99} Joncas, G. 1999, in proceedings of the 2nd
  Guillermo Haro Conference, Interstellar Turbulence, ed.\ J.~Franco
  \& A.~Carraminana, Cambridge University Press, p.154
\bibitem[Kolmogorov(1941)]{k41} Kolmogorov, A.~N., 1941, Dokl.\ Akad.\
  Nauk SSSR, 30, 301
\bibitem[Korpi et al.(1999)]{kbs99} Korpi, M.~J., Brandenburg, A.,
  Shukurov, A., Tuominen, I., \& Nordlund, \AA\ 1999, \apj, 514, 99
\bibitem[Larson(1979)]{l79} Larson, S. M. 1979, \mnras, 186, 479
\bibitem[Lazio et al.(1990)]{lsc90} Lazio, T. J., Spangler, S. R., \&
  Cordes, J. M. 1990, \apj, 363, 515
\bibitem[Mac Low \& Klessen(2004)]{mk04} Mac Low, M.-M., \& Klessen,
  R.~S. 2004, Rev.\ Mod.\ Phys., in press (astro-ph/0301093)
\bibitem[Maron \& Goldreich(2001)]{mg01} Maron, J., \& Goldreich,
  P. 2001, \apj, 554, 1175
\bibitem[McClure-Griffiths et al.(2001)]{mgd01} McClure-Griffiths,
  N.~M., Green, A.~J., Dickey, J.~M., Gaensler, B.~M., Haynes, R.~F.,
  \& Wieringa, M.~H 2001, \apj, 551, 394
\bibitem[Miller \& Scalo(1979)]{ms79} Miller, G.~E., \& Scalo,
  J.~M. 1979, \apjs, 41, 513
\bibitem[Minter \& Spangler(1996)]{ms96} Minter, A. H., \& Spangler,
  S. R. 1996, \apj, 458, 194
\bibitem[Norman \& Ferrara(1996)]{nf96} Norman, C.~A., \& Ferrara,
  A. 1996, \apj, 467, 280 
\bibitem[O'Dell(1991)]{o91} O'Dell, C.~R. 1991, in proceedings of the
  147th Symposium of the IAU, ed.\ E~Falgarone, F.~Boulanger, \&
  G.~Duvert, Kluwer Academic Publishers, Dordrecht, p.476
\bibitem[Padoan et al.(2001)]{pkg01} Padoan, P., Kim, S., Goodman, A.,
  \& Staveley-Smith, L. 2001, \apj, 555, 33
\bibitem[Paladini et al.(2003)]{pdd03} Paladini, R., Davies, R.~D.,
  \& DeZotti, G. 2003, preprint (astro-ph/0309350)
\bibitem[Prentice \& Ter Haar(1969)]{pt69} Prentice, A. J. R., \&
  Ter Haar, D. 1969, \mnras, 146, 423
\bibitem[Pynzar' \& Shishov(1999)]{ps99} Pynzar', A.~V., \& Shishov,
  V.~I. 1999, A.~Rep., 43, 7, 436
\bibitem[Rao \& Ananthakrishnan(1984)]{ra84} Rao, A.~P., \&
  Ananthakrishnan, S. 1984, \nat, 312, 707
\bibitem[Sault \& Killeen(2003)]{sk03} Sault, R. J., \& Killeen,
  N.~E.~B. 2003, The Miriad User's Guide (Sydney: Australia Telescope
  National Facility)
\bibitem[Sellwood \& Balbus(1999)]{sb99} Sellwood, J.~A., \& Balbus,
  S.~A. 1999, \apj, 511, 660
\bibitem[Simonetti \& Cordes(1986)]{sc86} Simonetti, J. H., \&
  Cordes, J. M. 1986, \apj, 310, 160
\bibitem[Simonetti et al.(1984)]{scs84} Simonetti, J.~H., Cordes,
  J.~M., \& Spangler, S.~R. 1984, \apj, 284, 126
\bibitem[Sokoloff et al.(1998)]{sbs98} Sokoloff, D.~D., Bykov, A.~A.,
  Shukurov, A., Berkhuijsen, E.~M., Beck, R., \& Poezd, A.~D. 1998,
  \mnras, 299, 189  
\bibitem[Spangler \& Reynolds(1990)]{sr90} Spangler, S. R., \&
  Reynolds, R. J. 1990, \apj, 361, 116
\bibitem[Spitzer(1978)]{s78} Spitzer, L. 1978, ``Physical processes in
  the interstellar medium'', New York Wiley-Interscience 
\bibitem[Stinebring \& Condon(1990)]{sc90} Stinebring, D.~R., \&
  Condon, J.~J. 1990, \apj, 352, 207
\bibitem[Sun \& Han(2004)]{sh04} Sun, X.~H, \& Han, J.~L. 2004, in
  ``The Magnetized Interstellar Medium'', ed.\ B.~Uyan\i ker, W.~Reich
  \& R.~Wielebinski, Copernicus GmbH, in press (astro-ph/0402180)
\bibitem[Vollmer \& Beckert(2002)]{vb02} Vollmer, B., \& Beckert,
  T. 2002, \aap, 382, 872

\end{thebibliography}
\end{document}